\documentclass[preprint,12pt,authoryear]{elsarticle}

\usepackage{amsmath, graphicx, mathtools, amssymb, booktabs}
\usepackage{multirow}
\usepackage{subcaption}
\usepackage{url}
\usepackage{natbib}

\begin{document}

\begin{frontmatter}
\author[label1]{Ivan Svetunkov}
\ead{i.svetunkov@lancaster.ac.uk}
\cortext[cor1]{Correspondance: I. Svetunkov, Centre for Marketing Analytics and Forecasting, Lancaster University Management School, Lancaster, Lancashire, LA1 4YX, UK.}
\address[label1]{Centre for Marketing Analytics and Forecasting,\\Management Science Department, Lancaster University, UK}

\title{Smooth forecasting with the \texttt{smooth} package in R}

\begin{abstract}
  There are many forecasting related packages in R with varied popularity, the most famous of all being \texttt{forecast}, which implements several important forecasting approaches, such as ARIMA, ETS, TBATS and others. However, the main issue with the existing functionality is the lack of flexibility for research purposes, when it comes to modifying the implemented models. The R package \texttt{smooth} introduces a new approach to univariate forecasting, implementing ETS and ARIMA models in Single Source of Error (SSOE) state space form and implementing an advanced functionality for experiments and time series analysis. It builds upon the SSOE model and extends it by including explanatory variables, multiple frequencies, and introducing advanced forecasting instruments. In this paper, we explain the philosophy behind the package and show how the main functions work.
\end{abstract}
\begin{keyword}
{forecasting, exponential smoothing, ets, arima, adam, \texttt{R}}
\end{keyword}

\end{frontmatter}





\section[Introduction]{Introduction}
R \citep{RStats}, being one of the most popular programming languages in academia, has many forecasting-related packages, implementing a variety of approaches. Among the well known ones, is the \texttt{forecast} package \citep{Hyndman2008}, which implements classical statistical forecasting models, such as ETS \citep[Error, Trend, Seasonality model based on the single source of error state space framework underlying exponential smoothing,][]{Hyndman2002, Hyndman2008b}, Theta \citep{Assimakopoulos2000}, TBATS \citep{DeLivera2011} and others. Some of these functions have also been implemented in \texttt{fable} package \citep{RFable}. It also implements the \texttt{auto.arima()} function for automatic selection of ARIMA orders. There are other packages implementing ARIMA, including \texttt{stats} \citep{RStats}, \texttt{robustarima} \citep{RRobustarima}, \texttt{tfarima} \citep{RTfarima} and \texttt{fable} \citep{RFable}. All these packages implement ready-to-use functions for specific situations and have been proven to work very well, but they do not have flexibility necessary for research purposes in the area of dynamic models and do not present a holistic approach to univariate forecasting models.

In order to address these issues, back in 2016, I developed a \texttt{smooth} package, which implemented models in the Single Source of Error framework and introduced flexibility allowing to conduct advanced experiments in the area of univariate forecasting models (e.g. using advanced losses, introducing explanatory variables, changing structures of models etc). This paper explains the main idea behind the \texttt{smooth} functions, summarises what they are created for and how to use them in forecasting and analytics.

\section[Single Source of Error]{Single Source of Error framework}
The main model underlying the \texttt{smooth} functions is explained in detail in \cite{SvetunkovAdam} monograph. It builds upon \cite{Hyndman2008b} model. Here we summarise only the main ideas. We start with the most popular pure additive model, underlying the majority of functions of \texttt{smooth} package \citep{RSmooth}. This is formulated as:
\begin{equation} \label{eq:ETSADAMStateSpacePureAdditive}
  \begin{aligned}
    {y}_{t} = &\mathbf{w}^\prime \mathbf{v}_{t-\mathbf{l}} + \epsilon_t \\
    \mathbf{v}_{t} = &\mathbf{F} \mathbf{v}_{t-\mathbf{l}} + \mathbf{g} \epsilon_t
  \end{aligned},
\end{equation}
where $\mathbf{w}$ is the measurement vector, $\mathbf{F}$ is the transition matrix, $\mathbf{g}$ is the persistence vector, $\mathbf{v}_{t-\mathbf{l}}$ is the vector of lagged components and $\mathbf{l}$ is the vector of lags, defining how each of the components of $\mathbf{v}_t$ needs to be shifted in time. Unlike the conventional state space model of \cite{Hyndman2008b}, the one implemented in \texttt{smooth} relies on lagged components rather than their transition. In the conventional case, the vector of states $\mathbf{v}_t$ always depends on the value of the vector on the previous observation, where the transition of signal happens from one component to another according to the matrix $\mathbf{F}$. Both the conventional and the proposed frameworks underlie exactly the same ETS and ARIMA models, but our approach simplifies calculations. For example, consider the ETS(A,A,A) model, which is written as \citep{Hyndman2008b}:
\begin{equation} \label{eq:ETSADAMAAA}
  \begin{aligned}
    y_{t} = & l_{t-1} & + & b_{t-1} & + & s_{t-m} & + & \epsilon_t \\
    l_t = & l_{t-1} & + & b_{t-1} & & & + & \alpha \epsilon_t \\
    b_t = & & & b_{t-1} & & & + & \beta \epsilon_t \\
    s_t = & & & & & s_{t-m} & + & \gamma \epsilon_t 
  \end{aligned}.
\end{equation}
where $y_t$ is the actual value, $l_{t-1}$ is the level, $b_{t-1}$ is the trend, $s_{t-m}$ is the seasonal component with periodicity $m$ (e.g. 12 for months of year data, implying that something is repeated every 12 months), $\alpha$, $\beta$ and $\gamma$ are the smoothing parameters and $\epsilon_t$ is an i.i.d. error term. According to \eqref{eq:ETSADAMStateSpacePureAdditive}, the model \eqref{eq:ETSADAMAAA} can be written as:
\begin{equation} \label{eq:ETSADAMAAAMatrixForm}
  \begin{aligned}
    y_t & = \begin{pmatrix} 1 & 1 & 1 \end{pmatrix}
    \begin{pmatrix}
      l_{t-1} \\ b_{t-1} \\ s_{t-m}
    \end{pmatrix} +
    \epsilon_t \\
    \begin{pmatrix}
      l_t \\ b_t \\ s_t
    \end{pmatrix} & =
    \begin{pmatrix}
      1 & 1 & 0 \\ 0 & 1 & 0 \\ 0 & 0 & 1
    \end{pmatrix}
    \begin{pmatrix}
      l_{t-1} \\ b_{t-1} \\ s_{t-m}
    \end{pmatrix} +
    \begin{pmatrix}
      \alpha \\ \beta \\ \gamma
    \end{pmatrix}
    \epsilon_t ,
  \end{aligned}
\end{equation}
while in the \citep{Hyndman2008b} form it would be:
\begin{equation} \label{eq:ETSADAMAAAConventional}
  \begin{aligned}
    y_t & = \begin{pmatrix} 1 & 1 & 1 & 0 & \dots & 0 \end{pmatrix}
    \begin{pmatrix}
      l_{t-1} \\ b_{t-1} \\ s_{1,t-1} \\ s_{2,t-1} \\ \vdots \\ s_{m,t-1}
    \end{pmatrix} +
    \epsilon_t \\
    \begin{pmatrix}
      l_{t} \\ b_{t} \\ s_{1,t} \\ \vdots \\ s_{m,t}
    \end{pmatrix} & =
    \begin{pmatrix}
      1 & 1 & \mathbf{0}^\prime_{m-1} & 0 \\ 0 & 1 & \mathbf{0}^\prime_{m-1} & 0 \\ 0 & 0 & \mathbf{0}^\prime_{m-1} & 1 \\ \mathbf{0}_{m-1} & \mathbf{0}_{m-1} & \mathbf{I}_{m-1} & \mathbf{0}_{m-1}
    \end{pmatrix}
    \begin{pmatrix}
      l_{t-1} \\ b_{t-1} \\ s_{1,t-1} \\ \vdots \\ s_{m,t-1}
    \end{pmatrix} +
    \begin{pmatrix}
      \alpha \\ \beta \\ \gamma \\ 0 \\ \vdots \\ 0
    \end{pmatrix}
    \epsilon_t .
  \end{aligned}
\end{equation}
Because of the size of matrices in \eqref{eq:ETSADAMAAAConventional} and the recursive nature of the model, applying it on seasonal data with $m$ higher than 24 becomes computationally expensive due to the multiplication of large matrices. The problem becomes even more serious when a model with multiple seasonal components is needed \citep[see, for example,][]{Taylor2003a}, because it then introduces several seasonal indices, increasing to the size of matrices in \eqref{eq:ETSADAMAAAConventional} even further. This issue is resolved in \eqref{eq:ETSADAMStateSpacePureAdditive}, because the introduction of additional components leads to increase of dimensionality proportional to the number of added components. Note that any extension of the conventional state space model results in the increase of its dimensionality, inevitably leading to computational difficulties. This is why the proposed model \eqref{eq:ETSADAMStateSpacePureAdditive} is more viable, flexible and this is why it was used in the development of \texttt{smooth} functions.

There is also a more general state space form, covering not only pure additive, but also pure multiplicative and mixed models. This is discussed in Chapter 4 of \cite{Hyndman2008b} and in Section 7.1 of \cite{SvetunkovAdam}. We do not discuss this model here but only point out that the multiplicative and mixed ETS models implemented in \texttt{smooth} are based on it.

Furthermore, \cite{Snyder1985} showed that ARIMA model can also be written in the SSOE state space form, this was then discussed in more detail in Chapter 11 of \cite{Hyndman2008b} and afterwards used by \cite{Svetunkov2019} to implement ARIMA in (\texttt{ssarima()} and \texttt{auto.ssarima()} functions from the \texttt{smooth} package) and apply it in supply chain context. Building upon that, \cite{SvetunkovAdam} implemented ARIMA (Chapter 9) in the state space form \eqref{eq:ETSADAMStateSpacePureAdditive}. So, the proposed framework does not stop on ETS models, it can be extended, for example, to include a combination of ETS+ARIMA.

The final piece of the puzzle is the regression model. It can also be represented in the SSOE state space form \eqref{eq:ETSADAMStateSpacePureAdditive}, as, for example, discussed in \cite{Koehler2012}. This is discussed in Chapter 9 of \cite{Hyndman2008b} and Chapter 10 of \cite{SvetunkovAdam}.

All of this means that the state space model \eqref{eq:ETSADAMStateSpacePureAdditive} presents a unified framework for working with ETS, ARIMA, regression and any of their combinations. This is implemented in \texttt{adam()} function of \texttt{smooth} package \citep{RSmooth}, supporting the following functionality:
\begin{enumerate}
    \item ETS;
    \item ARIMA;
    \item Regression;
    \item Time-varying parameters regression;
    \item Combination of (1), (2) and either (3), or (4);
    \item Automatic selection/combination of states of ETS;
    \item Automatic order selection for ARIMA;
    \item Variables selection for regression;
    \item Normal and non-normal distributions of the error term;
    \item Automatic selection of the most suitable distribution;
    \item Multiple seasonality;
    \item Occurrence part of the model to handle zeroes in data (in case of intermittent demand);
    \item Modelling scale of distribution \citep[GARCH-style models, see for example,][]{Engle1982};
    \item Handling uncertainty of estimates of parameters;
    \item Forecasting using any of the elements above.
\end{enumerate}
All these topics are covered in \cite{SvetunkovAdam}, so we will not focus on \texttt{adam} function here. However, there are special cases of the model \eqref{eq:ETSADAMStateSpacePureAdditive}, implementing specific functionality in the \texttt{smooth} package. We discuss the most important of them in this paper.

\section{Time series decomposition}
\begin{figure}[!htb]
	\centering
	\includegraphics[width=1\textwidth]{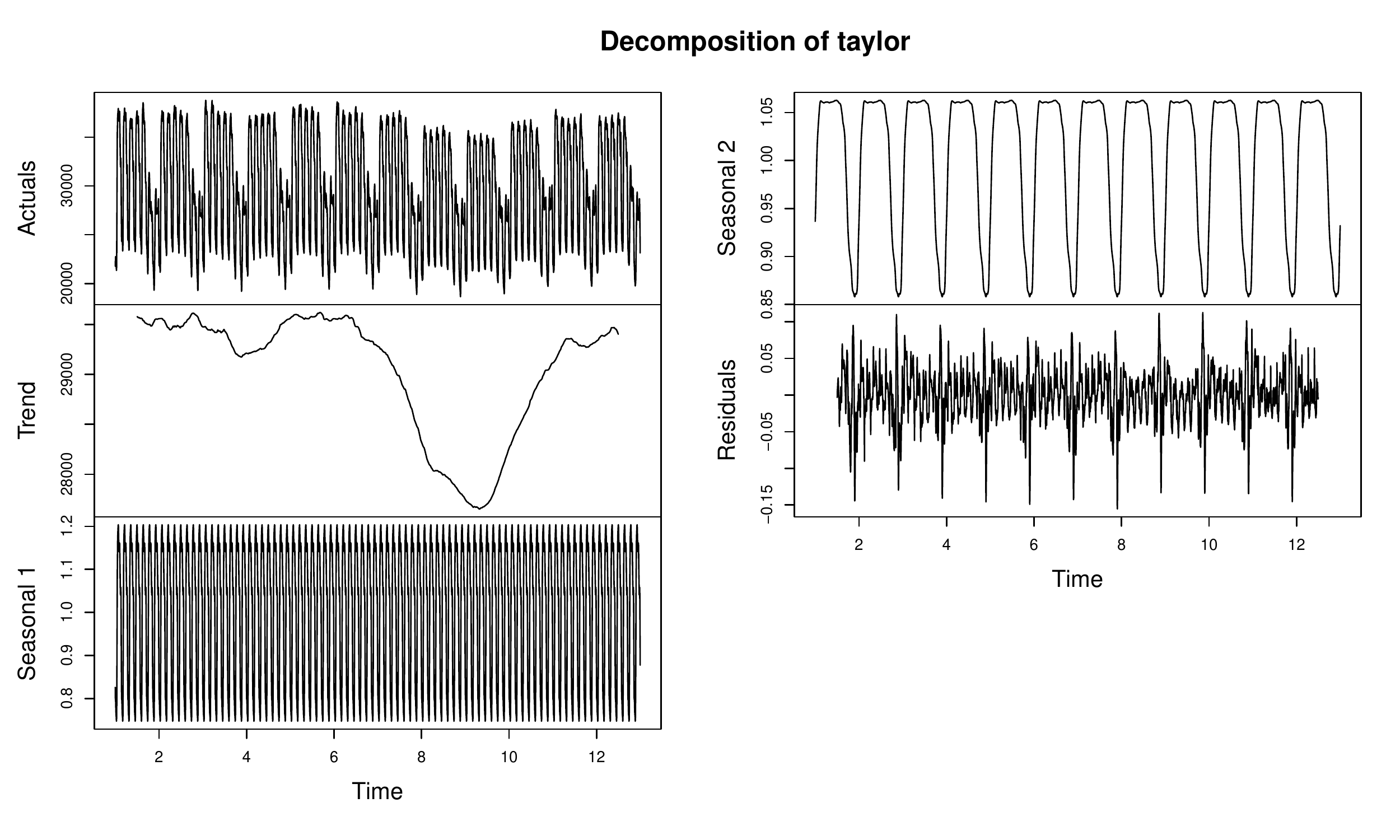}
	\caption{Decomposition of multiple seasonal time series according to \texttt{msdecompose()} function.}
	\label{fig:taylorDecomp}
\end{figure}

While \texttt{stats} package already implements the classical decomposition function, I have created a new one, which would handle the multiple seasonal data. This is called \texttt{msdecompose()}. It has exactly the same logic as the classical decomposition \citep{Persons1919}, but can be applied to the data with multiple seasonal cycles. A user can define what cycles there are in the data by setting the parameter \texttt{lags}, and choosing the type of seasonality via the parameter \texttt{type}. For example, \texttt{msdecompose()} can be applied to half-hourly electricity demand data \texttt{taylor} from \texttt{forecast} package in the following way:
\begin{verbatim}
    taylorDecomp <- msdecompose(taylor, lags=c(48,336),
                                type="m")
\end{verbatim}
which will result in an object that can be used for further analysis. Producing a plot from it would generate several figures (see documentation for \texttt{plot.smooth()} for details), the most interesting of which, the plot of time series components, is shown in Figure \ref{fig:taylorDecomp}.

The classical decomposition does not typically produce clear components -- the residuals in Figure \ref{fig:taylorDecomp} demonstrate the presence of seasonality because the approach assumes that the seasonal components are constant. Nonetheless, it can be a starting point for time series analysis. In \texttt{smooth}, it is used for the initialisation of states of ETS in case of seasonal data and is mainly needed when working with multiple seasonal data. However, if a researcher is interested in forecasting with seasonal decomposition, the function produces the object of the class \texttt{smooth} that supports the \texttt{forecast()} method, producing forecasts for the trend component of the decomposed data and then reconstructing the series based on the estimated seasonal components.

\section{Exponential smoothing} \label{sec:ETS}
Exponential smoothing is one of the most popular forecasting methods used in demand planning \citep{LCFweller2012}. As mentioned earlier, ETS underlies all exponential smoothing methods, and it is considered as an academic standard in forecasting \citep[it has been used in all the major forecasting competitions over the years, including][]{Makridakis2000, Athanasopoulos2011, Makridakis2020, Makridakis2022}. The conventional ETS model, developed by \cite{Hyndman2008b} assumes that the error term follows normal distribution. It is implemented in functions \texttt{ets()} from the \texttt{forecast} package and \texttt{ETS()} from the \texttt{fable}. Their counterpart in \texttt{smooth} package is called \texttt{es()}, but it is based on the state space model \eqref{eq:ETSADAMStateSpacePureAdditive} rather then the conventional one. Furthermore, while \texttt{ets()} supports only 15 ETS models, the \texttt{es()} implements all the theoretically possible 30 ETS models. The function also support fine tuning of parameters of model, allowing setting smoothing parameters values via \texttt{persistence} variable, initial values via \texttt{initial} and seasonal indices via \texttt{initialSeason} and pre-defining the values of parameters for the optimisation via the \texttt{B} parameter. Furthermore, the function supports explanatory variables via the \texttt{xreg} parameter, similar to how it is done in \texttt{arima} function from the \texttt{stats} package, allowing tuning the coefficients for regressors via \texttt{initialX} and selecting the most appropriate ones based on information criteria via the \cite{Sagaert2022} algorithm applied to residuals of the ETS model using the \texttt{regressors} parameter.

In terms of the ETS components selection, the mechanism used by default in \texttt{es()} can be summarised in the following steps:
\begin{enumerate}
    \item Apply ETS(A,N,N) to the data, calculate an information criterion (IC);
    \item Apply ETS(A,N,A) to the data, calculate IC. If it is lower than (1), then this means that there is some seasonal component in the data, move to step (3). Otherwise, go to (4);
    \item Apply ETS(M,N,M) model and calculate IC. If it is lower than the previous one, then the data exhibits multiplicative seasonality. Go to (4);
    \item Fit the model with the additive trend component and the seasonal component selected from the previous steps, which can be either ``N'', ``A'', or ``M''. Calculate IC for the new model and compare it with the best IC so far. If it is lower than the criteria of previously applied models, then there is a trend component in the data. If it is not, then the trend component is not needed;
    \item Form the pool of models based on steps (1) - (4), apply models and select the one with the lowest IC.
\end{enumerate}

This approach for components selection can be called branch-and-bound because instead of going through all possible models, it considers branches of models. For example, if there is no seasonality, then the respective component can be set to ``N'', thus removing the branch of seasonal models and reducing the pool of models to test from 30 to only 10 (including models already tested on the four steps).

Similarly to any other \texttt{smooth} function, \texttt{es()} supports several methods, including \texttt{plot()} for visual diagnostics of model and \texttt{forecast()} for forecasting. To demonstrate how they work, we apply \texttt{es()} to \texttt{AirPassengers} data from the \texttt{datasets} package.
\begin{verbatim}
    AirPassengersETS <- es(AirPassengers, h=12, holdout=TRUE)
\end{verbatim}
In the code above, we have specified the forecast horizon of 12 steps ahead and asked to exclude the last 12 observations from the training of the model, thus creating a test set (holdout) to see how the model performs in that part of the data. We can do diagnostics of the model in order to see if it has any obvious issues that could be resolved:
\begin{verbatim}
    par(mfcol=c(2,2))
    plot(AirPassengersETS, c(1,2,4,6))
\end{verbatim}

\begin{figure}[!htb]
	\centering
	\includegraphics[width=0.9\textwidth]{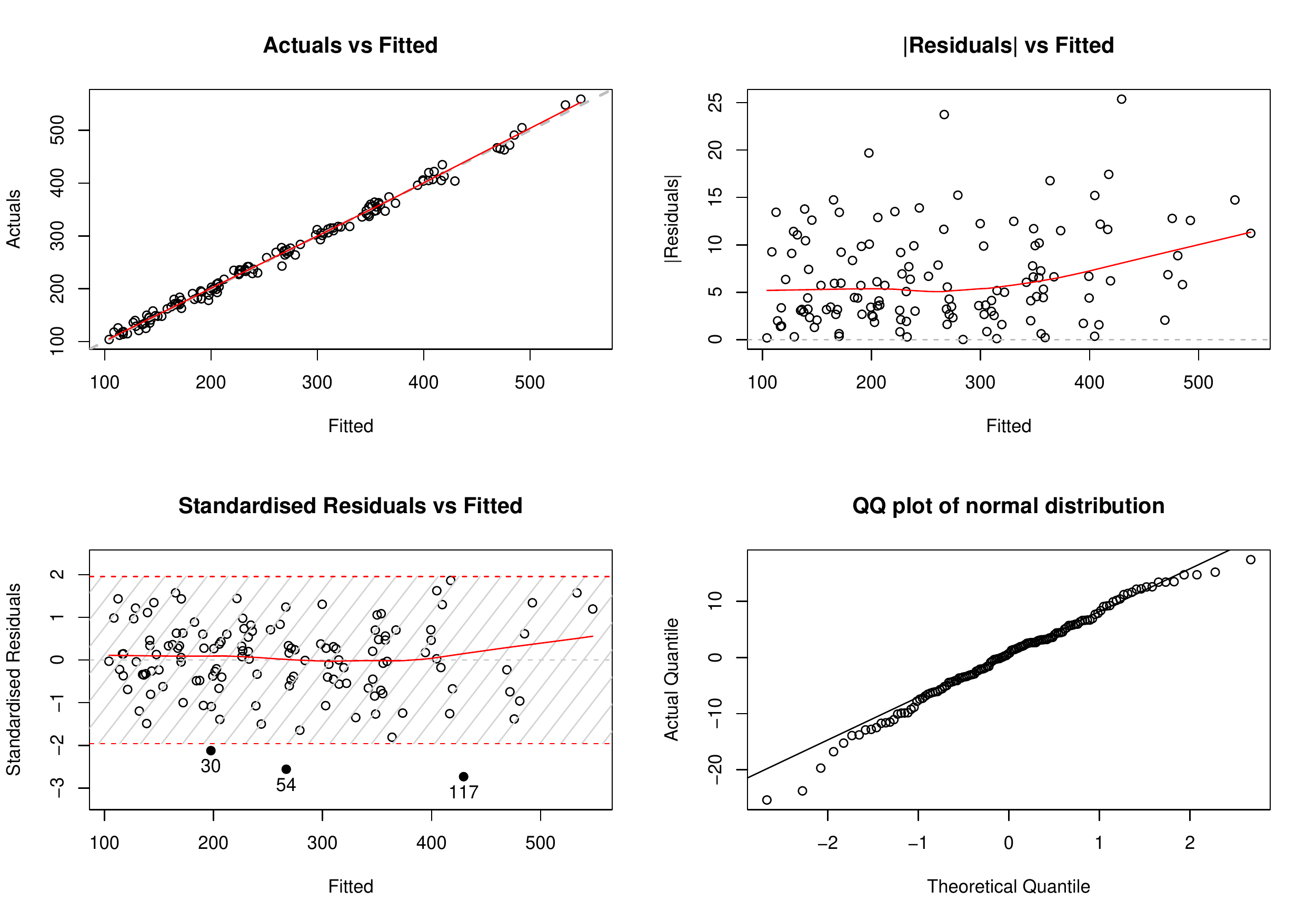}
	\caption{Diagnostics plots for ETS(A,M,M) model selected automatically on \texttt{AirPassengers} data by \texttt{es()} function.}
	\label{fig:AirPassengersETS}
\end{figure}

\begin{figure}[!htb]
	\centering
	\includegraphics[width=0.75\textwidth]{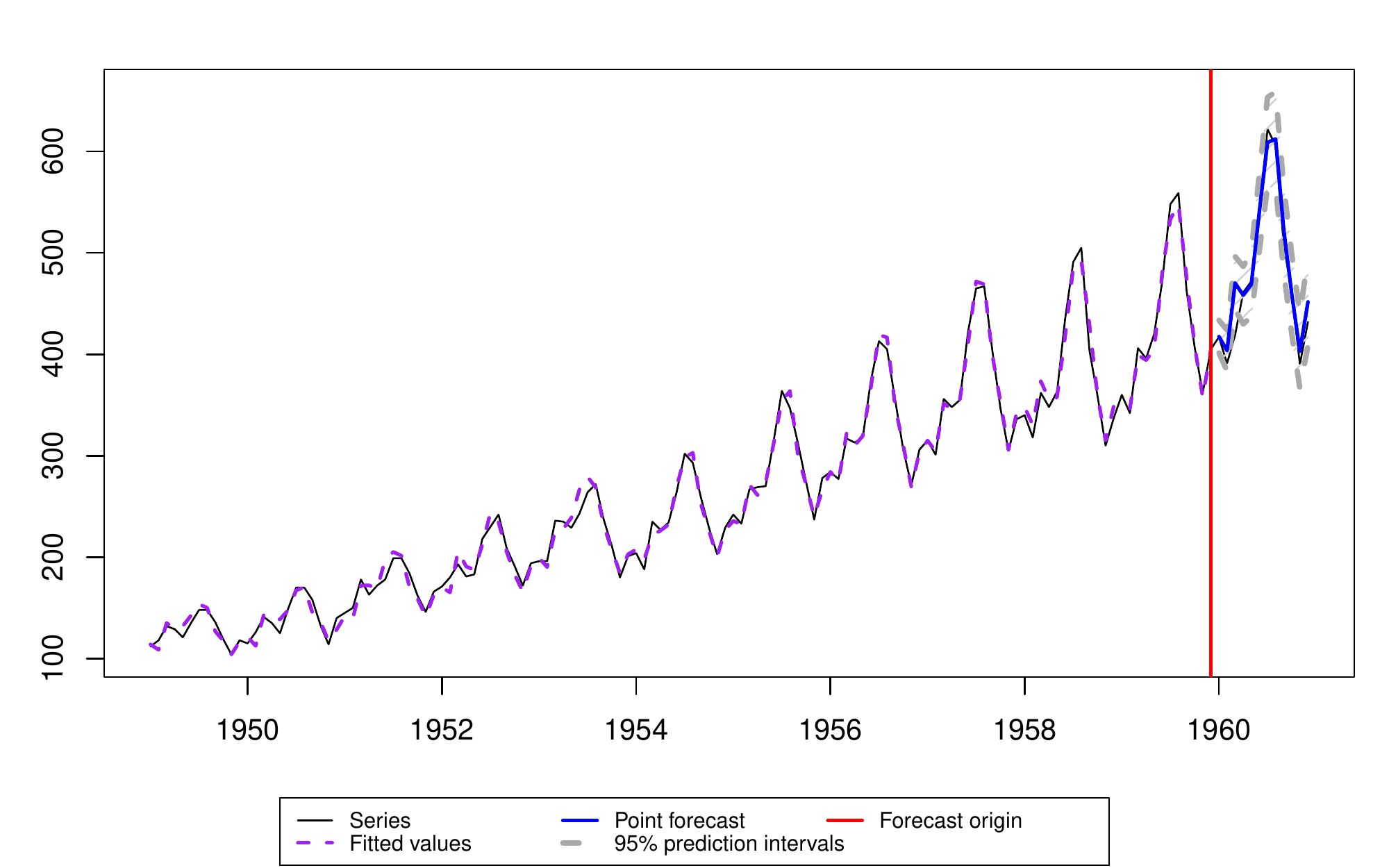}
	\caption{Forecast for \texttt{AirPassengers} data produced by \texttt{es()} function.}
	\label{fig:AirPassengersETSForecast}
\end{figure}

The resulting plot is shown in Figure \ref{fig:AirPassengersETS}. We do not aim to resolve the issues of the model in this paper, we merely demonstrate what can be done using \texttt{smooth} functions. The plots allow analysing the residuals for the possible issues related to heteroscedasticity, autocorrelation, outliers, wrong specification etc. Fixing the issues can be done by including explanatory variables and/or changing the transformations used in the model. After fixing the potential issues a researcher can produce forecasts from the estimated model, which is done using the \texttt{forecast()} method from \texttt{generics} package \citep{RGenerics}. But unlike the forecasting for the \texttt{ets()} and \texttt{ETS()}, the one from \texttt{smooth} supports several options, allowing choosing between a variety of prediction intervals (see documentation of \texttt{forecast.smooth()} method), allowing to produce one-sided interval (which is useful in case of pure multiplicative models on low-volume data, where the lower bound is typically equal to zero) and generating cumulative forecasts (which is useful in case of safety stock calculation in inventory management). We will use the default values of parameters, producing the parametric prediction interval:
\begin{verbatim}
    plot(forecast(AirPassengersETS, h=12))
\end{verbatim}
The code above will result in the plot in Figure \ref{fig:AirPassengersETSForecast}.

Figure \ref{fig:AirPassengersETSForecast} shows how the selected model fits the data, what point forecast it produces (solid bold blue line in the holdout part) and what prediction intervals it generated (a grey area in the holdout).

Continuing the theme of exponential smoothing, \texttt{smooth} also implements Complex Exponential Smoothing of \cite{Svetunkov2015} via the \texttt{ces()} function, which has the functionality similar to \texttt{es()} and supports the same set of methods.

Finally, as mentioned earlier, \texttt{adam()} implements ETS model as well and supports much more functionality. The main difference between the default ETS in \texttt{adam()} and  \texttt{es()} is that the former supports distributions other than normal and, by default, uses Gamma distribution in case of multiplicative error models.

\section{ARIMA}
Another important model, which is used often in forecasting, is ARIMA \citep{Box1976}. There are several functions implementing ARIMA in SSOE state space form in the \texttt{smooth} package.

The \texttt{ssarima()} (State Space ARIMA) function implements a state space ARIMA in the form discussed in Chapter 11 of \cite{Hyndman2008b}. The function that implements the order selection for State Space ARIMA is called \texttt{auto.ssarima()}. It does not rely on any statistical tests and selects orders based on information criteria. Both the model and the selection mechanism are explained in \cite{Svetunkov2019}.

The \texttt{msarima()} (Multiple Seasonal ARIMA) function relies on the state space model \eqref{eq:ETSADAMStateSpacePureAdditive}, introducing lagged components and thus substantially reducing the size of the transition matrix. This allows applying large multiple seasonal ARIMA models to the data. A thing to note is that because of this, the transition matrix, measurement, and state vectors of this model are formed differently than in \cite{Hyndman2008b}. In a general case, they are \citep[][Chapter 9]{SvetunkovAdam}:
\begin{equation} \label{eq:ADAMARIMAMatrices}
  \begin{aligned}
    \mathbf{F} = \begin{pmatrix} \eta_1 & \eta_1 & \dots & \eta_1 \\ \eta_2 & \eta_2 & \dots & \eta_2 \\ \vdots & \vdots & \ddots & \vdots \\ \eta_K & \eta_K & \dots & \eta_K \end{pmatrix}, & \mathbf{w} = \begin{pmatrix} 1 \\ 1 \\ \vdots \\ 1 \end{pmatrix}, \\
    \mathbf{g} = \begin{pmatrix} \eta_1 + \theta_1 \\ \eta_2 + \theta_2 \\ \vdots \\ \eta_K + \theta_K \end{pmatrix}, & \mathbf{v}_{t} = \begin{pmatrix} v_{1,t} \\ v_{2,t} \\ \vdots \\ v_{K,t} \end{pmatrix}, & \mathbf{l} = \begin{pmatrix} 1 \\ 2 \\ \vdots \\ K \end{pmatrix}
  \end{aligned} ,
\end{equation}
where $\eta_j$ is $j^{\text{th}}$ polynomial for the ARI part of the model, $\theta_j$ is the $j^{\text{th}}$ MA parameter and $K$ is the number of ARI/MA polynomials (whichever is the highest). To better understand how this model is formulated, consider an example of ARIMA(1,1,2), which can be written as:
\begin{equation} \label{eq:ARIMA(112)}
    (1 -\phi_1 B) (1-B) y_t = (1+\theta_1 B + \theta_2 B^2) \epsilon_t ,
\end{equation}
where $B$ is the backshift operator. This model can be written in the state space form \citep[see Chapter 9 of][for derivations]{SvetunkovAdam}:
\begin{equation} \label{eq:ADAMARIMA(112)}
    \begin{aligned}
    & y_{t} = v_{1,t-1} + v_{2,t-2} + \epsilon_t \\
    & v_{1,t} = (1+\phi_1) (v_{1,t-1} + v_{2,t-2}) + (1+\phi_1+\theta_1) \epsilon_t \\
    & v_{2,t} = -\phi_1 (v_{j,t-j} + v_{2,t-2}) + (-\phi_1+\theta_2) \epsilon_t
    \end{aligned} .
\end{equation}
In order to see that the model \eqref{eq:ADAMARIMA(112)} can be represented in the form \eqref{eq:ETSADAMStateSpacePureAdditive}, we need to set the following matrices and vectors:
\begin{equation} \label{eq:ADAMARIMAMatricesExample}
  \begin{aligned}
    \mathbf{F} = \begin{pmatrix} 1+\phi_1 & 1+\phi_1 \\ -\phi_1 & -\phi_1 \end{pmatrix}, & \mathbf{w} = \begin{pmatrix} 1 \\ 1 \end{pmatrix}, \\
    \mathbf{g} = \begin{pmatrix} 1+\phi_1+\theta_1 \\ -\phi_1+\theta_2 \end{pmatrix}, & \mathbf{v}_{t} = \begin{pmatrix} v_{1,t} \\ v_{2,t} \end{pmatrix}, & \mathbf{l} = \begin{pmatrix} 1 \\ 2 \end{pmatrix}
  \end{aligned}.
\end{equation}

Finally, as mentioned earlier, \texttt{adam()} function supports ARIMA\footnote{Note that in order to switch off the ETS part of the model in \texttt{adam()}, a user needs to specify \texttt{model="NNN"}.} as well, in the same form as \texttt{msarima()}. All the three functions have similar syntax for ARIMA, where a user needs to defined seasonal lags of model via \texttt{lags} vector, listing all seasonal frequencies that a model should have, and orders of model via \texttt{orders} variable, which in general accepts a named list of the style \texttt{orders=list(ar=c(1,2,3),i=c(1,2,3),ma=c(1,2,3))}, defining the order of AR, I and MA parts of the model for the respective lags. The ARIMA orders are designed this way to allow researchers to introduce as many lags as they need, supporting, for example, double and triple seasonal ARIMA. Note that due to its formulation \texttt{ssarima()} cannot handle high-frequency data and will slow down with the increase of the seasonal lag $m$.

\begin{figure}[!htb]
	\centering
	\includegraphics[width=0.75\textwidth]{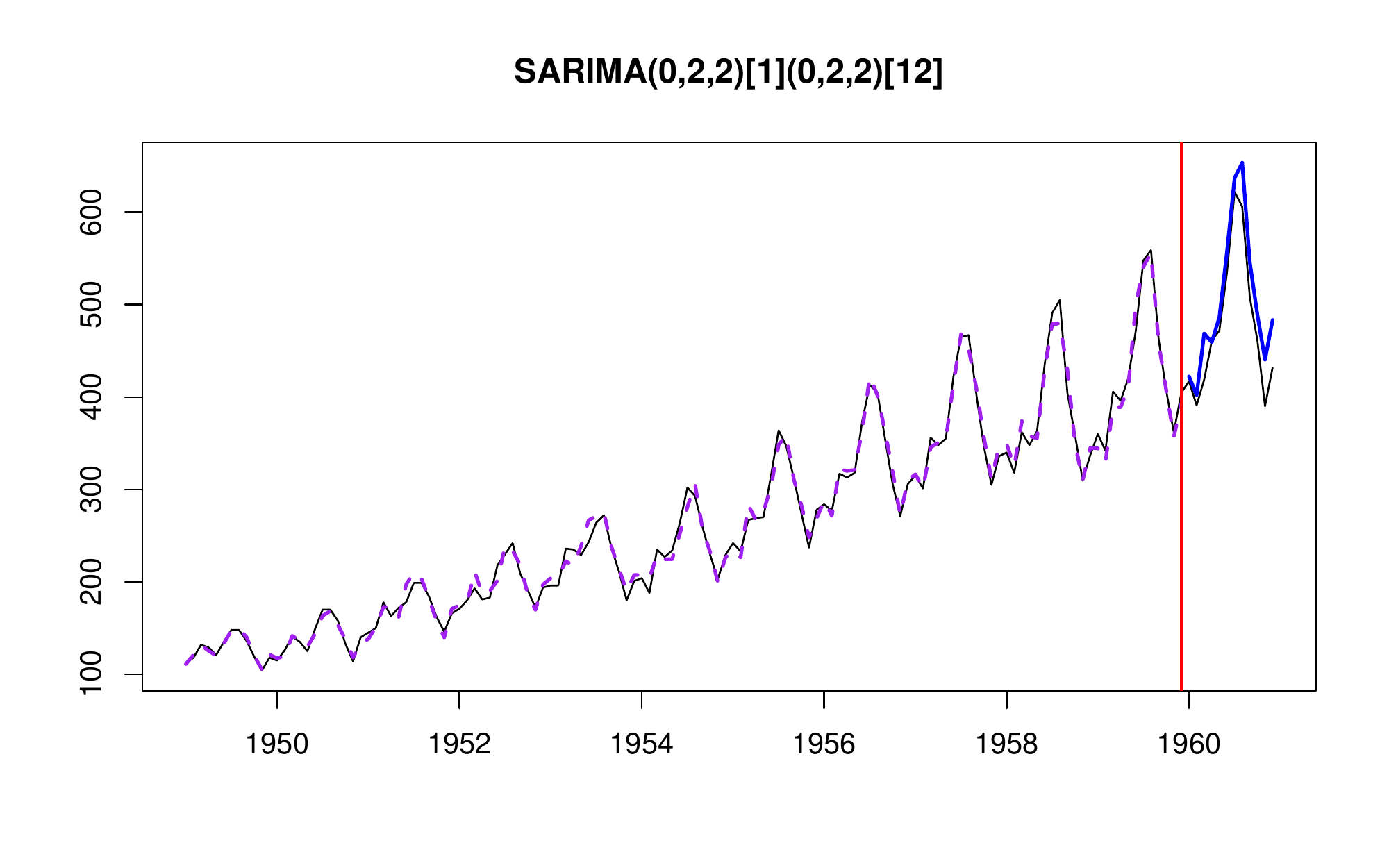}
	\caption{Forecast for \texttt{AirPassengers} data produced by \texttt{msarima()} function.}
	\label{fig:AirPassengersARIMA}
\end{figure}

Here is an example of a user defined SARIMA(0,2,2)(0,2,2)$_{12}$ model applied to the same \texttt{AirPassengers} data:
\begin{verbatim}
    AirPassengersARIMA <- msarima(AirPassengers, lags=c(1,12),
                                  orders=list(i=c(2,2),ma=c(2,2)),
                                  h=12, holdout=TRUE)
\end{verbatim}
In order to see how the model fits the data we can use the \texttt{plot} function, specifying \texttt{which=7}:
\begin{verbatim}
    plot(AirPassengersARIMA,7)
\end{verbatim}
after which we will get the plot shown in Figure \ref{fig:AirPassengersARIMA}.

Furthermore, all the \texttt{smooth} functions support one of the three mechanisms of initialisation:
\begin{enumerate}
    \item Optimisation -- the initial values of the state vector are estimated during the optimisation stage;
    \item Backcasting -- the initial values are produced via applying the model with optimised parameters to the reversed data, going recursively from the last observation to the first one;
    \item Manual -- the initials are provided by a user.
\end{enumerate}
These are regulated via the \texttt{initial} parameter in the functions. In the case of ARIMA, given the complexity of the task, \texttt{initial="backcasting"} typically works faster and more efficiently than the other two approaches.

If a researcher needs to have an ARIMA model with automatically selected orders, they can use \texttt{auto.ssarima()}, \texttt{auto.msarima()}, which will do that minimising the selected information criterion using the procedure described in \cite{Svetunkov2019} and in Section 15.2 of \cite{SvetunkovAdam}. In case of \texttt{adam()}, the automatic selection mechanism is switched on via addition of variable \texttt{select=TRUE} in the list for the \texttt{orders} parameter.

ARIMA models produced using the three functions above supports all the methods available for other \texttt{smooth} functions, including \texttt{plot()}, \texttt{actuals()}, \texttt{fitted()}, \texttt{residuals()} and \texttt{forecast()}.

\section{Simulation functions}
Another important set of functions supported by the \texttt{smooth} package is the simulation functions. They allow generating data from an assumed model. There are several functions in the package:

\begin{enumerate}
    \item \texttt{sim.es()} allows generating data from the selected ETS model with defined \texttt{persistence}, \texttt{initial} and \texttt{initialSeason} parameters;
    \item \texttt{sim.ces()} generates data from Complex Exponential Smoothing DGP;
    \item \texttt{sim.ssarima()} generates data from ARIMA model, allowing defining order of the model, AR, MA parameters and the value of the constant term (either intercept or drift, depending on the order of differences).
\end{enumerate}
If the parameters are not specified, they will be picked at random. All the functions above support a variety of distributions for the error term, allowing also to apply manually created ones. Here is an example of how to do the latter:
\begin{verbatim}
    customFunction <- function(n, mu, sd){
        return(log(abs(rnorm(n, mu, sd))));
    }
    x <- sim.es("ANN", obs=100,
                randomizer="customFunction", mu=0, sd=1)
\end{verbatim}
The simulation functions allow generating as many series as needed, which is regulated via \texttt{nsim} parameter.

\begin{figure}[!htb]
	\centering
	\includegraphics[width=0.75\textwidth]{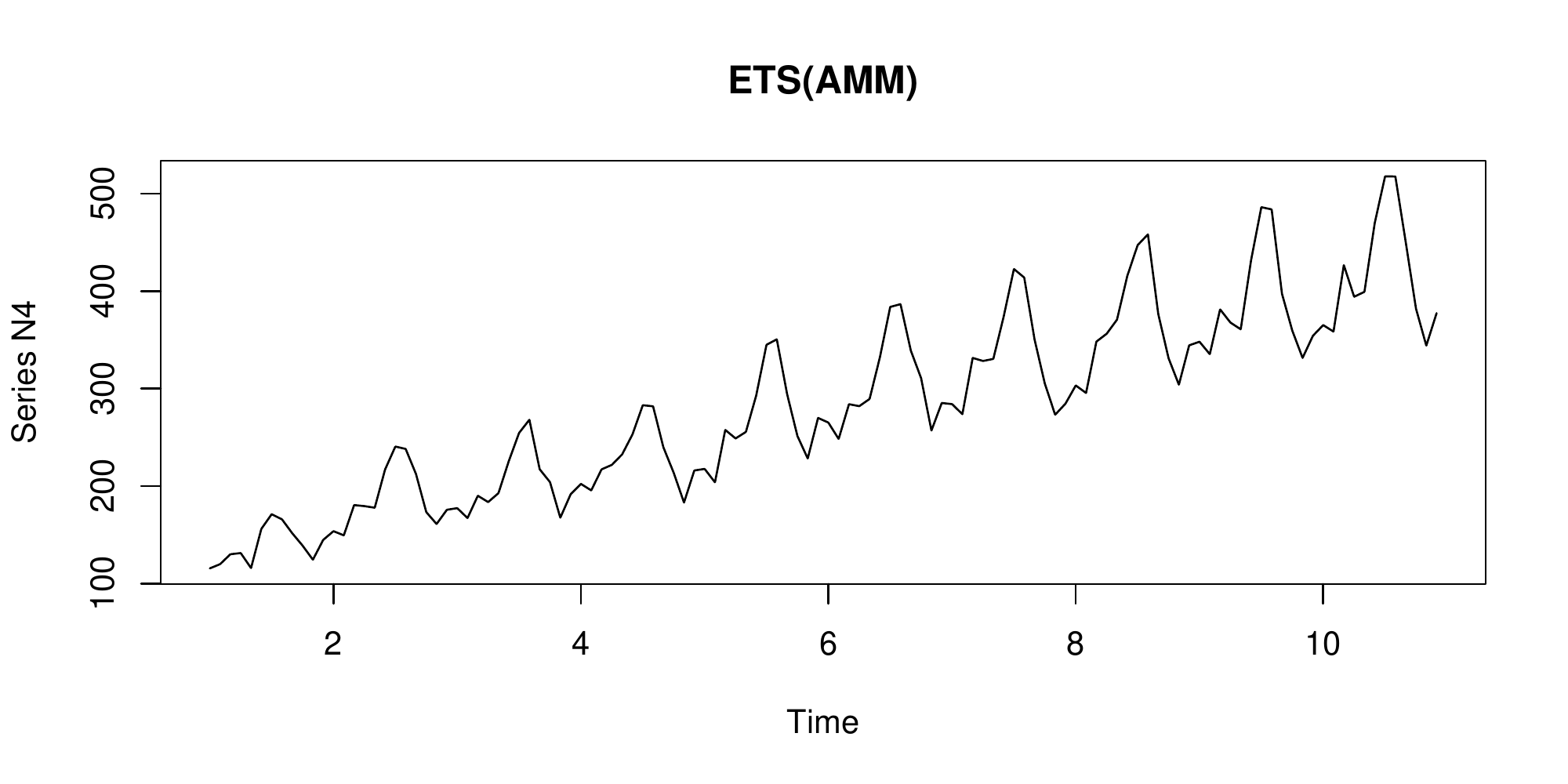}
	\caption{Simulated data from the \texttt{AirPassengersETS} model.}
	\label{fig:AirPassengersETSSimulate}
\end{figure}

Finally, the package also implements \texttt{simulate()} method, which extracts the parameters from the already estimated model to generate simulated data from it. In order to see how it works, we generate data from the \texttt{AirPassengersETS} model, estimated in Section \ref{sec:ETS}:
\begin{verbatim}
    x <- simulate(AirPassengersETS, obs=120, nsim=5)
    plot(x)
\end{verbatim}
The code above will generate five time series, and each one of them would look similar to the one shown in Figure \ref{fig:AirPassengersETSSimulate}.

As can be seen from the plot in Figure \ref{fig:AirPassengersETSSimulate}, the generated time series exhibits behaviour similar to the original time series. It even has a similar seasonal shape, but it has a different trend, increasing slower than in the original data.

\section{Other functions}
There are several other functions implemented in the package that are outside of the scope of this paper. Nonetheless, two of them are worth mentioning.

There is a Simple Moving Averages (SMA) function, \texttt{sma()}, implemented in state space model \eqref{eq:ETSADAMStateSpacePureAdditive}. This is based on the paper of \cite{Svetunkov2017} who showed that SMA(p) has an underlying AR(p) process with parameters restricted to $\phi_j=\frac{1}{p}$ for all $j=1, \dots, p$. The function also supports automatic order selection via information criteria as discussed in the original paper.

Another important function is \texttt{oes()}, which implements the occurrence part of model in case of intermittent demand. This is discussed in detail in Chapter 13 of \cite{SvetunkovAdam} and is based on \cite{Svetunkov2019a}.

Last but not least, \texttt{smooth} package has extensive vignettes with examples of application of almost all functions. It is available, for instance, on CRAN: \url{https://cran.r-project.org/package=smooth}.

\section[Benchmarking of functions]{Benchmarking of \texttt{smooth} functions}
Finally, to demonstrate how the \texttt{smooth} functions work, we conduct an experiment on M1 \citep{Makridakis1982}, M3 \citep{Makridakis2000} and Tourism  \citep{Athanasopoulos2011} competition data, where we evaluate seven models:

\begin{enumerate}
    \item ADAM ETS -- ETS model estimated via \texttt{adam()} function;
    \item ADAM ARIMA -- ARIMA model estimated via \texttt{adam()} function. We set \texttt{model="NNN"} to switch off ETS part of the model and use the following command to set the maximum ARIMA order to check: \\ \texttt{order=list(ar=c(3,2), i=c(2,1), ma=c(3,2), select=TRUE)};
    \item ES -- ETS model implemented in \texttt{es()} function, which is just a wrapper of \texttt{adam()};
    \item SSARIMA -- State Space ARIMA model estimated via \\ \texttt{auto.ssarima()};
    \item CES -- Complex Exponential Smoothing implemented in \texttt{auto.ces()} function from \texttt{smooth} package;
    \item ETS -- ETS model implemented in \texttt{ets()} function from \texttt{forecast} package;
    \item ARIMA -- ARIMA selected using \texttt{auto.arima()} function from \\ \texttt{forecast} package.
\end{enumerate}
We do not include \texttt{msarima()} in the experiment because the datasets under consideration do not have multiple seasonal time series. We have used the default values of parameters in all the functions. The forecasts were produced for each time series in the datasets for the horizons used in the original competitions to the part of the data not visible to the models. We produced point forecasts and 95\% prediction intervals to that part of series and evaluated the performance of models using the following measures:

\begin{itemize}
    \item MASE -- Mean Absolute Scaled Error by \cite{Hyndman2006};
    \item RMSSE -- Root Mean Scaled Squared Error introduced in \cite{Makridakis2022};
    \item Coverage -- percentage of observations in the holdout lying in the produced 95\% prediction interval;
    \item sMIS -- scaled Mean Interval Score from \cite{Makridakis2022};
    \item Time -- computational time in seconds spent for estimation and forecasts generation for each series.
\end{itemize}
For MASE, RMSSE, sMIS and time, the lower the value is, the better it is. For the coverage, the closer the value is to the nominal 95\%, the better it is.

The results of this experiment are summarised in Table \ref{tab:competitions}. Note that they might vary from one run to another because forecasts from some of the functions rely on simulations.

\begin{table}[!htb]
\centering
\begin{tabular}{l ccccc}
  \toprule
 & MASE & RMSSE & Coverage & sMIS & Time \\ 
  \midrule
ADAM ETS & \textbf{2.222} & \textbf{1.935} & \textit{0.885} & \textbf{2.122} & \textit{0.386} \\ 
  ES & \textit{2.224} & \textit{1.939} & \textbf{0.898} & \textit{2.196} & 0.477 \\ 
  CES & 2.271 & 1.958 & 0.812 & 3.465 & \textbf{0.236} \\ 
  ETS & 2.263 & 1.970 & 0.882 & 2.258 & 0.409 \\ 
  ARIMA & 2.300 & 1.987 & 0.834 & 3.007 & 1.425 \\ 
  ADAM ARIMA & 2.371 & 2.048 & 0.843 & 3.126 & 1.376 \\ 
  SSARIMA & 2.480 & 2.133 & 0.802 & 3.356 & 1.811 \\ 
   \bottomrule
\end{tabular}
   \caption{Error measures for each of the model evaluated on M1, M3 and Tourism competitions, aggregated using mean values. The \textbf{boldface} indicates the best performing models, while the \textit{italic} indicates the second best ones.}
   \label{tab:competitions}
\end{table}

As can be seen from Table \ref{tab:competitions}, ADAM ETS outperforms all other models in terms of MASE, RMSSE and sMIS, although the difference between it and other ETS implementations does not look substantial. Note that it works slightly slower than CES. The ETS from \texttt{forecast} package performs slightly worse than the \texttt{smooth} implementations on these datasets. Comparing ARIMA implementations, the one from \texttt{auto.arima()} is more accurate and faster than ADAM ARIMA and SSARIMA, although it was not able to beat the ETS models. Note however that ARIMA produces lower coverage than ADAM ARIMA does and works slower.

This example demonstrates that the developed functions work efficiently and can be applied to a wide variety of time series. Table \ref{tab:competitions} summarises an overall aggregate performance, which does not mean that the winning models always perform the best. Their performance will vary from one series to another, and in some instances, the models that performed poorly in this experiment would perform much better \citep[for example, SSARIMA performed very well on supply chain data with a short history as discussed in][]{Svetunkov2019}.

\section[Conclusions]{Conclusions}
In this paper, I have discussed the philosophy behind the models implemented in the \texttt{smooth} package for R. The state space model used in the functions differs from the conventional one, allowing to introduce more components and using more complex models efficiently. We have discussed how ETS and ARIMA are implemented in this framework and what an analyst can achieve with them. Finally, we have demonstrated how the models implemented in the \texttt{smooth} functions perform on an example of M1, M3 and Tourism competitions data.

This paper merely introduced the framework, the models and the functions. As mentioned earlier, the main idea of the \texttt{smooth} functions is to give a researcher flexibility. A reader interested in learning more about the framework is advised to read the online monograph of \cite{SvetunkovAdam} and to study examples in the vignettes of the \texttt{smooth} package in R \citep{RSmooth}.

\bibliographystyle{elsarticle-harv}
\bibliography{library}

\begin{thebibliography}{30}
\expandafter\ifx\csname natexlab\endcsname\relax\def\natexlab#1{#1}\fi
\expandafter\ifx\csname url\endcsname\relax
  \def\url#1{\texttt{#1}}\fi
\expandafter\ifx\csname urlprefix\endcsname\relax\def\urlprefix{URL }\fi

\bibitem[{Assimakopoulos and Nikolopoulos(2000)}]{Assimakopoulos2000}
Assimakopoulos, V., Nikolopoulos, K., 2000. {The theta model: a decomposition
  approach to forecasting.} International Journal of Forecasting 16, 521--530.

\bibitem[{Athanasopoulos et~al.(2011)Athanasopoulos, Hyndman, Song, and
  Wu}]{Athanasopoulos2011}
Athanasopoulos, G., Hyndman, R.~J., Song, H., Wu, D.~C., 2011. {The tourism
  forecasting competition}. International Journal of Forecasting 27~(3),
  822--844.

\bibitem[{Box and Jenkins(1976)}]{Box1976}
Box, G., Jenkins, G., 1976. {Time series analysis: forecasting and control}.
  Holden-day, Oakland, California.

\bibitem[{{De Livera} et~al.(2011){De Livera}, Hyndman, and
  Snyder}]{DeLivera2011}
{De Livera}, A.~M., Hyndman, R.~J., Snyder, R.~D., 2011. {Forecasting Time
  Series With Complex Seasonal Patterns Using Exponential Smoothing}. Journal
  of the American Statistical Association 106~(496), 1513--1527.

\bibitem[{Engle(1982)}]{Engle1982}
Engle, R.~F., jul 1982. {Autoregressive Conditional Heteroscedasticity with
  Estimates of the Variance of United Kingdom Inflation}. Econometrica 50~(4),
  987.

\bibitem[{Gallego(2021)}]{RTfarima}
Gallego, J.~L., 2021. tfarima: Transfer Function and ARIMA Models. R package
  version 0.2.1.
\newline\urlprefix\url{https://CRAN.R-project.org/package=tfarima}

\bibitem[{Hyndman and Khandakar(2008)}]{Hyndman2008}
Hyndman, R.~J., Khandakar, Y., 2008. Automatic time series forecasting: the
  forecast package for {R}. Journal of Statistical Software 26~(3), 1--22.

\bibitem[{Hyndman and Koehler(2006)}]{Hyndman2006}
Hyndman, R.~J., Koehler, A.~B., 2006. {Another look at measures of forecast
  accuracy}. International Journal of Forecasting 22~(4), 679--688.

\bibitem[{Hyndman et~al.(2008)Hyndman, Koehler, Ord, and Snyder}]{Hyndman2008b}
Hyndman, R.~J., Koehler, A.~B., Ord, J.~K., Snyder, R.~D., 2008. {Forecasting
  with Exponential Smoothing}. Springer Berlin Heidelberg.

\bibitem[{Hyndman et~al.(2002)Hyndman, Koehler, Snyder, and
  Grose}]{Hyndman2002}
Hyndman, R.~J., Koehler, A.~B., Snyder, R.~D., Grose, S., 2002. {A state space
  framework for automatic forecasting using exponential smoothing methods}.
  International Journal of Forecasting 18~(3), 439--454.

\bibitem[{Kaluzny and {TIBCO Software Inc.}(2021)}]{RRobustarima}
Kaluzny, S., {TIBCO Software Inc.}, 2021. robustarima: Robust ARIMA Modeling. R
  package version 0.2.6.
\newline\urlprefix\url{https://CRAN.R-project.org/package=robustarima}

\bibitem[{Koehler et~al.(2012)Koehler, Snyder, Ord, and Beaumont}]{Koehler2012}
Koehler, A.~B., Snyder, R.~D., Ord, J.~K., Beaumont, A., 2012. {A study of
  outliers in the exponential smoothing approach to forecasting}. International
  Journal of Forecasting 28~(2), 477--484.

\bibitem[{Makridakis et~al.(1982)Makridakis, Andersen, Carbone, Fildes, Hibon,
  Lewandowski, Newton, Parzen, and Winkler}]{Makridakis1982}
Makridakis, S., Andersen, A.~P., Carbone, R., Fildes, R., Hibon, M.,
  Lewandowski, R., Newton, J., Parzen, E., Winkler, R.~L., 1982. {The accuracy
  of extrapolation (time series) methods: Results of a forecasting
  competition}. Journal of Forecasting 1~(2), 111--153.

\bibitem[{Makridakis and Hibon(2000)}]{Makridakis2000}
Makridakis, S., Hibon, M., 2000. {The M3-Competition: results, conclusions and
  implications}. International Journal of Forecasting 16, 451--476.

\bibitem[{Makridakis et~al.(2020)Makridakis, Spiliotis, and
  Assimakopoulos}]{Makridakis2020}
Makridakis, S., Spiliotis, E., Assimakopoulos, V., 2020. {The M4 Competition:
  100,000 time series and 61 forecasting methods}. International Journal of
  Forecasting 36~(1), 54--74.

\bibitem[{Makridakis et~al.(2022)Makridakis, Spiliotis, and
  Assimakopoulos}]{Makridakis2022}
Makridakis, S., Spiliotis, E., Assimakopoulos, V., oct 2022. {M5 accuracy
  competition: Results, findings, and conclusions}. International Journal of
  Forecasting 38~(4), 1346--1364.

\bibitem[{O'Hara-Wild et~al.(2021)O'Hara-Wild, Hyndman, and Wang}]{RFable}
O'Hara-Wild, M., Hyndman, R., Wang, E., 2021. fable: Forecasting Models for
  Tidy Time Series. R package version 0.3.1.
\newline\urlprefix\url{https://CRAN.R-project.org/package=fable}

\bibitem[{{R Core Team}(2022)}]{RStats}
{R Core Team}, 2022. R: A Language and Environment for Statistical Computing. R
  Foundation for Statistical Computing, Vienna, Austria.
\newline\urlprefix\url{https://www.R-project.org/}

\bibitem[{Sagaert and Svetunkov(2022)}]{Sagaert2022}
Sagaert, Y., Svetunkov, I., 2022. {Trace Forward Stepwise: Automatic Selection
  of Variables in No Time}.

\bibitem[{Snyder(1985)}]{Snyder1985}
Snyder, R.~D., 1985. {Recursive Estimation of Dynamic Linear Models}. Journal
  of the Royal Statistical Society, Series B (Methodological) 47~(2), 272--276.

\bibitem[{Svetunkov(2022)}]{SvetunkovAdam}
Svetunkov, I., 2022. Forecasting and analytics with adam. Monograph.
  OpenForecast, (version: 2022-04-18).
\newline\urlprefix\url{https://openforecast.org/adam/}

\bibitem[{Svetunkov(2023)}]{RSmooth}
Svetunkov, I., 2023. smooth: Forecasting Using State Space Models. R package
  version 3.2.0.
\newline\urlprefix\url{https://github.com/config-i1/smooth}

\bibitem[{Svetunkov and Boylan(2019)}]{Svetunkov2019a}
Svetunkov, I., Boylan, J., 2019. {Multiplicative state-space models for
  intermittent time series}.

\bibitem[{Svetunkov and Boylan(2020)}]{Svetunkov2019}
Svetunkov, I., Boylan, J.~E., 2020. {State-space ARIMA for supply-chain
  forecasting}. International Journal of Production Research 58~(3), 818--827.

\bibitem[{Svetunkov et~al.(2022)Svetunkov, Kourentzes, and Ord}]{Svetunkov2015}
Svetunkov, I., Kourentzes, N., Ord, J.~K., 8 2022. Complex exponential
  smoothing. Naval Research Logistics (NRL), 31.

\bibitem[{Svetunkov and Petropoulos(2018)}]{Svetunkov2017}
Svetunkov, I., Petropoulos, F., 2018. {Old dog, new tricks: a modelling view of
  simple moving averages}. International Journal of Production Research
  56~(18), 6034--6047.

\bibitem[{Taylor(2003)}]{Taylor2003a}
Taylor, J.~W., 2003. {Short-term electricity demand forecasting using double
  seasonal exponential smoothing}. Journal of the Operational Research Society
  54~(8), 799--805.

\bibitem[{{Warren M. Persons}(1919)}]{Persons1919}
{Warren M. Persons}, 1919. {General Considerations and Assumptions}. The Review
  of Economics and Statistics 1~(1), 5--107.

\bibitem[{Weller and Crone(2012)}]{LCFweller2012}
Weller, M., Crone, S.~F., November 2012. Supply chain forecasting: Best
  practices \& benchmarking study. Tech. rep., Lancaster Centre for
  Forecasting.

\bibitem[{Wickham et~al.(2022)Wickham, Kuhn, and Vaughan}]{RGenerics}
Wickham, H., Kuhn, M., Vaughan, D., 2022. generics: Common S3 Generics not
  Provided by Base R Methods Related to Model Fitting. R package version 0.1.2.
\newline\urlprefix\url{https://CRAN.R-project.org/package=generics}

\end{thebibliography}

\end{document}